\documentclass[showpacs,prl,onecolumn,aps,superscriptaddress,preprintnumbers,nofootinbib]{revtex4}
\usepackage[T1]{fontenc}
\usepackage[latin9]{inputenc}
\usepackage{amsmath,amssymb}
\usepackage{epsfig}
\usepackage{graphicx}
\usepackage{mathrsfs}
\usepackage{amsmath}
\usepackage{amsfonts}
\usepackage{epstopdf}
\usepackage{color}
\def\slashchar#1{\setbox0=\hbox{$#1$}     		
   \dimen0=\wd0                                 	
   \setbox1=\hbox{/} \dimen1=\wd1               	
   \ifdim\dimen0>\dimen1                        	
      \rlap{\hbox to \dimen0{\hfil/\hfil}}      	
      #1                                        	
   \else                                        	
      \rlap{\hbox to \dimen1{\hfil$#1$\hfil}}   	
      /                                         	
   \fi}

\renewcommand{\vec}{\boldsymbol}
\newcommand{\be}{\begin{equation}}
\newcommand{\ee}{\end{equation}}
\newcommand{\bea}{\begin{eqnarray}}
\newcommand{\eea}{\end{eqnarray}}
\newcommand{\ba}{\begin{array}}
\newcommand{\ea}{\end{array}}

\def\eq#1{{Eq.~(\ref{#1})}}
\def\fig#1{{Fig.~\ref{#1}}}
\newcommand{\bas}{\bar{\alpha}_S}

\newcommand{\nn}{\nonumber}

\newcommand{\Lb}{\left(}
\newcommand{\Rb}{\right)}
\newcommand{\h}{\frac{1}{2}}

\begin{document}

\title{ Multiplicity distribution of dipoles in QCD from  Le, Mueller and Munier equation}

 \author{Eugene ~ Levin}
\email{leving@tauex.tau.ac.il, eugeny.levin@usm.cl}
\affiliation{Department of Particle Physics, School of Physics and Astronomy,
Raymond and Beverly Sackler
 Faculty of Exact Science, Tel Aviv University, Tel Aviv, 69978, Israel}\affiliation{ Departamento de F\'\i sica,
Universidad T$\acute{e}$cnica Federico Santa Mar\'\i a   and
Centro Cient\'\i fico-Tecnol$\acute{o}$gico de Valpara\'\i so,
Casilla 110-V,  Valparaiso, Chile}

\date{\today}

\pacs{13.60.Hb, 12.38.Cy}

\begin{abstract}

In this paper we derived in QCD  the BFKL linear, inhomogeneous equation for the factorial moments of  multiplicity distribution ($M_k$) from LMM  equation. In particular, the equation for the average multiplicity of the color-singlet dipoles ($N$)
turns out to be the homogeneous  BFKL  while  $M_k\,\,\propto\,\,N^k$ at small $x$. 
Second, using the diffusion approximation for the BFKL kernel we show that the factorial moments  are equal to: $
M_k\,\,=\,\,k!\ N \,\Lb N \,\,-\,\,1\Rb^{k-1}$ which leads to the multiplicity distribution:  $ \frac{\sigma_n}{\sigma_{ \rm in}}\,\,=\,\,\frac{1}{N}\,\Lb \frac{N\,-\,1}{N}\Rb^{n - 1}$. 
 We also suggest a procedure for finding corrections to this multiplicity distribution which  will be useful for descriptions of the experimental data.
 
 \end{abstract}
\maketitle

\vspace{-0.5cm}
\tableofcontents

\section{Introduction}

  During the past several years  a  robust 
relation
 between the principle features of high energy scattering and entanglement
 properties of the hadronic wave function   have been in focus of the high
 energy and nuclear physics communities  
\cite{KUT,PES,KOLU1,PESE,KHLE,BAKH,BFV,HHXY,KOV1,GOLE1,GOLE2,KOV2,NEWA,LIZA,FPV,TKU,KOV3,GOLE,KHLE1}.  In this paper, we continue to explore
 the relation between the entropy in the
 parton approach\cite{BJ,FEYN,BJP,Gribov} and  the entropy of
 entanglement in a proton wave function\cite{KHLE}.  In Ref.\cite{KHLE}, it is proposed that
 parton distributions can be defined in terms of the entropy of entanglement between the spatial region probed by deep inelastic scattering (DIS) and the rest of the proton.  This approach leads to a simple relation $S = \ln N $  between the average number of color-singlet dipoles and the entropy of the produced hadronic state $S$. This simple relation shows that  a proton becomes a maximally entangled state in the region of small Bjorken $x$.  All these conclusions were made from estimates in the simple, even naive  model for QCD cascade of color-singlet dipoles. However,  it has been demonstrated  in Refs.\cite{BAKH,GOLE1,GOLE2,TKU,KHLE1}  
 that these 
ideas are in qualitative and, partly, in  quantitative  agreement with the
 available experimental data.  Actually, it is shown in Ref.\cite{KHLE} that the simple cascade of color-singlet dipoles leads to the multiplicity distribution:

 \be \label{MULTD}
\frac{\sigma_n}{\sigma_{ \rm in}}\,\,=\,\,\frac{1}{N}\,\Lb \frac{N\,-\,1}{N}\Rb^{n - 1}\,
\ee
where $N$ is the average number of dipoles.

The goal of this paper is to study the multiplicity distribution and the
 entanglement entropy in the effective theory for  QCD at high energies
 (see Ref.\cite{KOLEB} for a general review). We have approached this problem in Refs.\cite{KHLE,GOLE} and have demonstrated that \eq{MULTD} arises in QCD cascades.  In    this paper we analyze the multiplicity distribution for Balitsky-Kovchegov (BK) cascade\cite{BK} in which one dipole at low energy generates a large number color-singlet dipoles at high energy. The equation for such a cascade  are known (see Refs.\cite{KOLEB,MUCD,LELU1,LELU2} ) and the first try to solve them have been undertaken in Ref.\cite{GOLE}. However, in this paper we return to this problem and study the multiplicity distribution using the new equation (Le, Mueller and Munier (LMM) equation)  for the probability generating function that has been derived in Ref.\cite{LMM}.
 
  In the next section we derive LMM equation from the equation for the BK parton cascade. In the rest of the paper we  discuss the equations for the factorial moments that follow from the LMM equation. We  show that every factorial moment satisfies the linear but inhomogeneous equation with the Balitsky, Fadin, Kuraev and Lipatov (BFKL) kernel\cite{BFKL,LI}. We  attempt  to solve these equations and  demonstrate that in the diffusion approximation to the BFKL kernel  factorial moments are equal to:
  
  \be \label{C1}
M_k\,\,=\,\,k!\ N \,\Big( N \,\,-\,\,1\Big)^{k-1}
\ee  
  
  We show that \eq{C1} leads to \eq{MULTD}.
  
  In section VI we suggest an approach  to go beyond    diffusion approximation, which cannot give a reliable description of the experimental data even in the leading order of perturbative QCD.  In this approach we propose  to solve exactly the    equations for the factorial moments and using the difference between the exact solution and  \eq{C1} ($\Delta M_k\,=\,M_k\Lb \rm exact\Rb \,-\,M_k\Lb \eq{C1}\Rb$) we develop  the way how to estimate the multiplicity distributions beyond diffusion approximation. 
   In conclusion section  we summarize our results.

\section{ General features of the  cascade of color-singlet dipoles in QCD }

In QCD at large number of colors $N_c$ ($N_c\,\,\gg\,\,1$) the color-singlet dipoles play the role of partons (see Ref.\cite{KOLEB} for review). 
 As discussed in Refs.\cite{KOLEB,MUCD,LELU1,LELU2}  for them we can write the following  equations:
  \bea  \label{PC1}
&&\frac{\partial\,P_n\left(Y, \vec{r }, \vec{b};\,\vec{r}_1,\vec{ b}_1,\,\vec{r}_2 , \vec{b}_2\dots \vec{r}_i ,\vec{b}_i,
\dots \vec{r}_n, \vec{b}_n \right)}{ 
\partial\, Y }\,=\,-\,
\sum^n_{i=1}\,\omega_G(r_i) \,
P_n\left(Y, \vec{r }, \vec{b};\,\vec{r}_1,\vec{ b}_1,\,\vec{r}_2 , \vec{b}_2\dots \vec{r}_i ,\vec{b}_i,
\dots \vec{r}_n, \vec{b}_n \right) \\
&&~~~~~~~~~~~~~~~~~~~~~~~~~~~~~~~~~~~~~~~~~~~~~+\,\,\bas\,\sum^{n-1}_{i=1} \,\frac{(\vec{r}_i\,+\, 
\vec{r}_n)^2}{(2\,\pi)\,r^2_i\,r^2_n}\,
P_{n - 1}\left(Y, \vec{r},\vec{b};\,\vec{r}_1,\vec{b}_1,
\dots  (\vec{r}_i \,+\, \vec{r}_n), \vec{b}_{in},\dots \vec{r}_{n-1},\vec{b}_n
 \right)\nn
\eea
  where $P_n\Lb Y ; \{r_i,b_i\}\Rb$ is the probability to have $n$-dipoles
 of size $r_i$,  at impact parameter $b_i$ and  at rapidity $Y$\footnote
{ In the lab. frame rapidity $Y$ is equal to   $Y = y_{\rm dipole ~r} \,-\,y_{\rm dipoles~r_i}$, where $y_{\rm dipole ~r}$ is the rapidity of the incoming fast dipole and $y_{\rm dipole~r_i}$ is the rapidity of dipoles $r_i$. Note, that all rapidities of dipoles $r_i$ are the same in  \eq{PC1}.}
  . $\vec{b}_{in} $ in \eq{PC1} is equal to $\vec{b}_{in}
 \,=\,\vec{b}_i \,+\,\h \vec{r}_i \,=\,\vec{b}_n \,-\,\h \vec{r}_i$. $\omega_G\Lb r\Rb$ is defined below in  \eq{OMG}.

  \eq{PC1} is a typical cascade equation in which the first term
 describes the reduction   of  the probability to find $n$ dipoles
 due to the possibility that one of $n$ dipoles can decay into two dipoles 
of
 arbitrary sizes, while the second term  describes  the growth due to the 
splitting
 of $(n-1)$ dipoles into $n$ dipoles.

The initial condition for the DIS scattering is
\be \label{PCIC}
P_1 \Lb Y  =  0,  \vec{r},\vec{b} ; \vec{r}_1,\vec{b}_1\Rb\,\, =\,\,\,\delta^{(2)}\Lb \vec{r}\,-\,\vec{r}_1\Rb\,\delta^{(2)}\Lb \vec{b}\,-\,\vec{b}_1\Rb;~~~~~~~P_{n>1}\Lb Y  = 0; \{r_i\}\Rb \,=\,0\ee
which corresponds to the fact that we are discussing a dipole of 
 definite size which develops the parton cascade.
Since $P_n\Lb Y ; \{r_i\}\Rb$ is the probability to find dipoles $\{r_i\}$,
 we have the following sum rule 

\be \label{SUMRU}
\sum_{n=1}^\infty\,\int \prod^n_{i=1} d^2 r_i \,d^2 b_i \,P_n\Lb Y ; \{\vec{r}_i\,\vec{b}_i\}\Rb\,\,=\,\,1 ,
\ee
i.e. the sum of all probabilities is equal to 1.

 This QCD  cascade leads to Balitsky-Kovchegov (BK) equation 
\cite{BK,KOLEB} for the  amplitude and gives the theoretical
 description of the DIS.  We introduce the generating functional\cite{MUCD}

\be \label{Z}
Z\Lb Y, \vec{r},\vec{b}; [u_i]\Rb\,\,=\,\,\sum^{\infty}_{n=1}\int P_n\Lb Y,\vec{r},\vec{b};\{\vec{r}_i\,\vec{b}_i\}\Rb \prod^{n}_{i=1} u\Lb \vec{r}_i\,\vec{b}_i\Rb\,d^2 r_i\,d^2 b_i
\ee
 where $u\Lb \vec{r}_i\,\vec{b}_i\Rb \equiv\,u_i$ is an arbitrary function.
 The initial conditions of \eq{PCIC}  and the sum rules of \eq{SUMRU} take 
the following form for the functional $Z$:
\begin{subequations}
\bea
Z\Lb Y=0, \vec{r},\vec{b}; [u_i]\Rb &\,\,=\,\,&u\Lb \vec{r},\vec{b}\Rb;\label{ZIC}\\
Z\Lb Y, r,[u_i=1]\Rb &=& 1; \label{ZSR}
\eea
\end{subequations}

Multiplying both parts of \eq{PC1} by $\prod^{n}_{i=1} u\Lb \vec{r}_i\,\vec{b}_i\Rb$ and integrating over $r_i$ and $b_i$ we obtain the following linear functional equation\cite{LELU2};
\begin{subequations}
\bea
&&\hspace{-0.7cm}\frac{\partial Z\Lb Y, \vec{r},\vec{b}; [u_i]\Rb}{\partial \,y} =\int d^2 r'\,  K\Lb \vec{r}',\vec{r} - \vec{r'}|\vec{r}\Rb\Bigg( - u\Lb r, b\Rb\,\,+\,\,u\Lb \vec{r}',\vec{b} + \h(\vec{r} - \vec{r}') \Rb \,u\Lb \vec{r} - \vec{r}',\vec{b} + \h\vec{r}'\Rb\Bigg) \frac{\delta\,Z}{\delta \,u\Lb r, b \Rb};\label{EQZ}\\
&&  K\Lb \vec{r}',\vec{r} - \vec{r'}|\vec{r}\Rb\,=\frac{1}{2 \,\pi}\frac{r^2}{r'^2\,(\vec{r} - \vec{r}')^2} ;\,~~~~~
 \omega_G\Lb r\Rb\,\,=\,\,\int d^2 r'  K\Lb \vec{r}',\vec{r} - \vec{r'}|\vec{r}\Rb; \label{OMG}
 \eea
\end{subequations}
where $y\,\,=\,\,\bas\,Y$.

Searching for the solution of the form     $Z\Lb [ u(r_i,b_i,Y)]\Rb$  
for the initial conditions of \eq{ZIC}, \eq{EQZ} can be re-written as
 the non-linear equation \cite{MUCD}:
\be \label{NEQZ}
\frac{\partial Z\Lb Y, \vec{r},\vec{b}; [u_i]\Rb}{\partial \,y}\,=\,\int d^2 r' K\Lb \vec{r}',\vec{r} - \vec{r'}|\vec{r}\Rb\Bigg\{Z\Lb r',  \vec{b} + \h(\vec{r} - \vec{r}'); [u_i]\Rb  \,Z\Lb \vec{r} - \vec{r'},  \vec{b} + \h\vec{r}'; [u_i]\Rb
\,\,-\,\,Z\Lb Y, \vec{r},\vec{b}; [u_i]\Rb\Bigg\}
\ee
Therefore, the QCD parton cascade of \eq{PC1} takes into account 
 non-linear evolution. 
\section{Derivation of  Le, Mueller and Munier (LMM) equation }
In this section we derive the LMM equation which is  proposed in Ref.\cite{LMM}. First, we introduce the same notations as in Ref.\cite{LMM}:
 \be \label{W}
 \tilde{w}_n\Lb r, b, y\Rb\,\,=\,\, \int \prod^n_{i=1} d^2 r_i \,d^2 b_i \,P_n\Lb Y ; \{\vec{r}_i\,\vec{b}_i\}\Rb
 \ee
One can see that $ \tilde{w}_n\Lb r, b, y\Rb$ is the probability that the dipole  with size $r$ produces $n$ dipoles with all possible sizes. \eq{ZSR} reads as 
\be\label{ZSRM}
 \sum_{n=1}^\infty \tilde{w}_n\Lb r, b, y\Rb\,\,=\,\,1
 \ee

Taking all $u_i\Lb r_i,b_i\Rb\,\,=\,\,\lambda$ one can see that we can re-write  \eq{Z} in the form:
\be \label{WLA} 
Z\Lb Y, \vec{r},\vec{b}; [u_i = \lambda]\Rb\,\,\equiv\,\tilde{w}_\lambda\Lb \vec{r}, \vec{b}, y\Rb\,\,=\,\, \sum_{n=1}^\infty  \lambda^n\,\,\tilde{w}_n\Lb \vec{r}, \vec{b}, y\Rb
\ee

Plugging \eq{WLA} into \eq{NEQZ} we obtain the LMM equation in the form:
\be \label{LMM}
\frac{\partial \tilde{w}_\lambda\Lb \vec{r}, \vec{b}, y\Rb}{\partial\,\,y}  =\int d^2 \,r' \, K\Lb \vec{r}',\vec{r} - \vec{r'}|\vec{r}\Rb\Bigg\{ \tilde{w}_\lambda\Lb \vec{r}', \vec{b} + \h \Lb \vec{r}\,-\,\vec{r}'\Rb, y\Rb\,\,\tilde{w}_\lambda\Lb \vec{r}\,-\,\vec{r}', b\,+\,\h \vec{r}', y\Rb\,\,-\,\,\tilde{w}_\lambda\Lb \vec{r},  \vec{b}, y\Rb\Bigg\}
\ee
  In addition, discussing multiplicity distribution $P_n \,=\,\frac{\sigma_n}{\sigma_{tot}}$, where $\sigma_n$ is the cross section for production of $n$  color-singlet dipoles, we need to integrate  $ w_n\Lb r, b, y\Rb$ over $b$. In this case the initial condition for the dipole cascade takes the form:
  \be \label{PCIC1}
P_1 \Lb Y  =  0,  \vec{r}; \vec{r}_1\Rb\,\, =\,\,\,\delta^{(2)}\Lb \vec{r}\,-\,\vec{r}_1\Rb\;~~~~~~~P_{n>1}\Lb Y  = 0; \{r_i\}\Rb \,=\,0
\ee  
 which leads to the probabilities, that do not depend on impact parameters.  Since in \eq{PC1} $b$  enters as a parameter,  $P_n\Lb  Y; r_i\Rb$, which does not depend on $b_i$, is also a solution to \eq{PC1},  which satisfies 
 \eq{PCIC1}.

  \eq{LMM} reduces to
 \be \label{LMM1}
\frac{\partial \tilde{w}_\lambda\Lb \vec{r},  y\Rb}{\partial\,\,y}  =\int d^2 \,r' \, K\Lb \vec{r}',\vec{r} - \vec{r'}|\vec{r}\Rb\Bigg\{ \tilde{w}_\lambda\Lb \vec{r}', y\Rb\,\,\tilde{w}_\lambda\Lb \vec{r}\,-\,\vec{r}', y\Rb\,\,-\,\,\tilde{w}_\lambda\Lb \vec{r}, y\Rb\Bigg\}
\ee 
 
 \eq{LMM1} is a particular case of the general equation that has been derived in Ref.\cite{LMM}.  In this paper 
instead of \eq{W} the more general form of this equation is proposed, viz.:
\be \label{WS}
 w_n\Lb r, b,  y,  y_0 \Rb\,\,=\,\, \int \prod^n_{i=1}\Big(  d^2 r_i \,d^2 b_i \,S\Lb r_i, b_i, y_0\Rb\Big) \,P_n\Lb Y ; \{\vec{r}_i\,\vec{b}_i\}\Rb
 \ee
where $S\Lb r_i, b_i, y_0\Rb$ is the scattering S-matrix for elastic interaction of the dipole with size $r_i$  at rapidity $Y_0 $($y_0 \,=\,\bas\,Y_0$)  and at impact parameter $b_i$  with the target at $Y=0$. Since $S$ is a unitarity matrix, $ w_n\Lb r, b, y; y_0\Rb$ is the probability that the dipole  with size $r$ produces $n$ dipoles with all possible sizes, which interact with the target.
Bearing this in mind,  we see that \eq{ZSRM} holds for  $w_\lambda\Lb r, b,  y; y_0\Rb$, which is defined as

\be \label{WLA} 
Z\Lb Y, \vec{r},\vec{b}; [u_i = \lambda \,S\Lb r_i, b_i, y_0\Rb]\Rb\,\,\equiv\,w_\lambda\Lb \vec{r}, \vec{b}, y, y_0\Rb\,\,=\,\, \sum_{n=1}^\infty  \lambda^n\,\,w_n\Lb \vec{r}, \vec{b}, y, y_0\Rb
\ee

For the case of $P_n$ which do not depend on $b_i$, 
inserting in \eq{NEQZ} $u_i\Lb r_i,b_i,\Rb \,\,=\,\,\lambda S\Lb r_i,b_i,y_0\Rb$  we see that we obtain the  LMM equation in its original form (see Ref.\cite{LMM}):

\be \label{LMMGEN}
\frac{\partial w_\lambda\Lb \vec{r}, b,  y, y_0\Rb}{\partial\,\,y}  = \int d^2 \,r' \, K\Lb \vec{r}',\vec{r} - \vec{r'}|\vec{r}\Rb\Bigg\{w_\lambda\Lb \vec{r}', y, y_0\Rb\,\,w_\lambda\Lb \vec{r}\,-\,\vec{r}', y, y_0\Rb\,\,-\,\,w_\lambda\Lb \vec{r},  y, y_0\Rb\Bigg\}
\ee

\section{Average number of color-singlet  dipoles}
The average number of dipoles can be calculated using the following formula:
\be \label{N}
N\Lb \vec{r},  y, y_0 \Rb\,\,=\,\,\langle| n |\rangle = \sum^\infty_{n=1} n\,w_n\Lb r, y, y_0\Rb\,=\,\frac{\partial\,\, w_\lambda\Lb \vec{r},  y, y_0\Rb}{\partial\,\,\lambda} \Big{|}_{\lambda\,=\,1}
\ee

Differentiating \eq{LMM} with respect to $\lambda$ we obtain that
\be \label{NEQN}
 \frac{\partial N\Lb \vec{r}, y , y_0\Rb}{\partial\,\,y}\,\,=\,\,\int d^2 r' K\Lb \vec{r}',\vec{r} - \vec{r'}|\vec{r}\Rb\Bigg\{N\Lb \vec{r}', y, y_0 \Rb \,+\,N\Lb \vec{r}\,-\,\vec{r}', y, y_0 \Rb \,\,-\,\,N\Lb \vec{r}, y, y_0 \Rb\Bigg\}
 \ee

 \eq{NEQN} shows that the average number of dipoles satisfies the linear  BFKL \cite{BFKL,LI} equation and 
 increases in the region of small $x$ (large $y$).  Therefore, we see that the general  QCD cascade  reproduces the main observation of Ref.\cite{KHLE} which was made in the oversimplified model  for the QCD cascade. In this model the dependence on the size of the dipoles were neglected. 
 
 The general solution takes the following form:

 \be \label{SOLN}
N\Lb \vec{r}, y, y_0 \Rb\,\,\,=\,\,\,\int\limits^{\epsilon \,+\,i\,\infty}_{\epsilon \,-\,i\,\infty}\!\!\!\! \frac{d \gamma}{2\,\pi\,i}\,e^{ \chi\Lb \gamma\Rb\,y \,\,+\,\,\gamma\,\xi}\,n_{in}\Lb \gamma, y_0\Rb
\ee
where  $\xi \,\,=\,\, \ln\Lb\frac{1}{ r^2}\Rb$  and $\chi(\gamma)$ is the BFKL kernel:

\bea \label{BFKLKER}
\chi\Lb \gamma \Rb  =  2 \psi\Lb 1\Rb \,-\,\psi\Lb \gamma\Rb\,-\,\psi\Lb 1 - \gamma\Rb
= \left\{\begin{array}{l}\,\,\,\frac{1}{\gamma}\,\,\,\mbox{for}\,\,\,\gamma\,\to\,0  \leftarrow \,\,\mbox{double log approximation(DLA)};\\ \\
\underbrace{ 4 \ln 2}_{\omega_0} \,\,+\,\, \underbrace{14 \zeta\Lb 3\Rb}_{D} \Lb \gamma - \h\Rb^2 \,\,\mbox{for}\,\,\,\gamma\,\to\,\h\, \leftarrow \,\,\mbox{diffusion approximation(DA)};\\  \end{array}
\right.
\eea

 where $\psi(z)$ is the Euler $\psi$-function (see Ref.\cite{RY} formula{ \bf 8.36}). 
 
 $n_{in}\Lb \gamma\Rb$ has to be found from the initial condition $  N\Lb \vec{r}, \vec{b}, y = 0,y_0=0 \Rb \,\,=\,\,1$ (see \eq{PCIC} and \eq{ZIC}). It gives
 \be \label{NIC}
 n_{in}\Lb \gamma\Rb\,\,=\,\,\frac{1}{\gamma}
 \ee
 
 Introducing  multiplicity in the momentum representation:
 \be \label{MMOM}
N\Lb k_T, y\Rb \,\,=\,\,\int d^2 r\,e^{- i \vec{k}_T \cdot \,\vec{r}} \frac{N\Lb r,y\Rb}{r^2},
\ee 
 we can re-write \eq{NEQN}  in the form:
 
 \be \label{NEQM}
\frac{\partial\,N\Lb \vec{k}_T,  y,  y_0 \Rb}{\partial\,y}\,\,=\,\, \int \frac{d^2 k_T}{(2 \pi)^2} K\Lb \vec{k}_T,\vec{k}'_T\Rb\,\,N\Lb \vec{k}'_T, y, y_0 \Rb 
\ee 

where $K\Lb \vec{k}_T,\vec{k}'_T\Rb$ is the BFKL kernel in momentum representation:

\be \label{KERM}
K\Lb \vec{k}_T,\vec{k}'_T\Rb\,\,N\Lb \vec{k}'_T, y, y_0 \Rb\,=\,\frac{1}{\Lb \vec{k}_T - \vec{k}'_T\Rb^2} \,\,N\Lb \vec{k}'_T, y, y_0 \Rb\,\,-\,\,\frac{k^2_T}{\Lb \vec{k}_T - \vec{k}'_T\Rb^2\,\Lb\Lb \vec{k}_T - \vec{k}'_T\Rb^2\,+\,k'^2_T\Rb }\,\,N\Lb \vec{k}_T, y, y_0 \Rb
\ee

 Solution to this equation has the same form of \eq{SOLN} but with the replacement of $\xi \,\to \xi' =    \ln k^2_T$ and
 \be \label{NINM}
 n_{in}\Lb \gamma\Rb\,\,=\,\,\,\chi\Lb \gamma\Rb/\gamma
 \ee
  \eq{NINM} reproduces  the value of $N\Lb k_T, y=0, y_0=0\Rb$, since \eq{MMOM} at $y=0$ leads to $ N\Lb k_T, y\Rb \,=\,\ln k^2_T + {\cal O}\Lb 1/k^2_T\Rb$. It is worth mentioning that  (1) in the double log approximation $ n_{in}\Lb \gamma\Rb = 1/\gamma^2$ , which leads to   $ N\Lb k_T, y\Rb \,=\,\ln k^2_T  $  and (2) in semi-classical approximation, which we will use below, we can neglect all corrections of the order of $1/k^2_T$.
 
\section{Equations for moments of the multiplicity distribution}
\subsection{The second moment}
\subsubsection{Equation}
We start  a derivation of the  evolution equation for the moments of the multiplicity distributions   considering the second moment, which has the following form:

\be \label{M2}
M_2\Lb \vec{r}, y,  y_0 \Rb\,\,=\,\,\,\langle| n (n\,-\,1) |\rangle = \sum^\infty_{n=1} n\,(n\,-\,1) \,w_n\Lb r, y, y_0\Rb\,=\,\frac{\partial^2\,\, w_\lambda\Lb \vec{r}, \ y, y_0\Rb}{\partial\,\,\lambda^2} \Big{|}_{\lambda\,=\,1}
\ee
 Taking the second derivative with respect to $\lambda$ from \eq{LMMGEN} we obtain the equation for $M_2\Lb \vec{r},  y,  y_0 \Rb$
 \bea \label{M2EQ}
\frac{\partial\,M_2\Lb \vec{r}, y,  y_0 \Rb}{\partial\,y}\,\,&=&\,\, \int d^2 r' K\Lb \vec{r}',\vec{r} - \vec{r'}|\vec{r}\Rb\Bigg\{M_2\Lb \vec{r}', y, y_0 \Rb \,+\,M_2\Lb \vec{r}\,-\,\vec{r}',  y, y_0 \Rb\nn\\
 & \,\,+&\,\,2\,N\Lb \vec{r}' , y, y_0 \Rb \,N\Lb \vec{r}\,-\,\vec{r}' , y ,y_0\Rb -\,\,M_2\Lb \vec{r},  y, y_0 \Rb\Bigg\} 
\eea

 \eq{M2EQ} is a linear but inhomogeneous  equation with the  inhomogeneous  term, which is determined by the multiplicity of the dipoles.
 
 The initial conditions (see \eq{PCIC} and \eq{ZIC}) for this equation is
  \be \label{M2IC}
  M_2\Lb \vec{r},  y\,=\,y_0 , y_0 \Rb  \,\,=\,\,0
  \ee

First, lets us start to solve \eq{M2EQ} making first iteration at small $y = \Delta y$.

For $y=0$ $N\Lb r, y =0\Rb \,=\,1$ and $M_2\Lb r, y =0\Rb=0$, and hence 
\be \label{M21I}
M^{(1)}_2 \,=\,2\,\Delta y \,\omega_G\Lb r \Rb
\ee
 where $\omega  $ is given by \eq{OMG}. The first iteration of \eq{NEQN} leads to
$N^{(1)} \,=\,1\,\,+\,\,\,\Delta y \,\omega_G\Lb r \Rb$. Comparing these two estimates one can see that the first iteration can be written as the expansion of the solution $ M_2\Lb r, y\Rb\,\,=\,\,2\Lb N^2\Lb r, y\Rb\,\,-\,\,
N\Lb r, y \Rb\Rb$ with respect to $\Delta \,y$. Hence, we see, that at small $\Delta y$ we obtain the simple expression for $M_2$, which turns out to be the same as for the multiplicity distribution of \eq{MULTD} for the simple toy model \cite{KHLE}.  We will try to prove this equation below, but we have  succeeded only in the  semi-classical  approximation.


\subsubsection{General solution}

The general solution to \eq{M2EQ} we can obtain going to momentum representation:

\be \label{M2MOM}
m_2\Lb k_T, y\Rb \,\,=\,\,\int\frac{d^2 r}{(2 \pi)^2}e^{- i \vec{k}_T \cdot \,\vec{r}} \frac{M_2\Lb r, y\Rb}{r^2}
\ee

\eq{M2EQ} takes the form:
\be \label{M2EQMOM}
\frac{\partial\,m_2\Lb \vec{k}_T,  y,  y_0 \Rb}{\partial\,y}\,\,=\,\, \int \frac{d^2 k_T}{(2 \pi)^2} K\Lb \vec{k}_T,\vec{k}'_T\Rb\,\,m_2\Lb \vec{k}'_T,  y, y_0 \Rb 
  \,\,+\,\,2\,N^2\Lb \vec{k}_T,  y, y_0 \Rb  
\ee

Taking the Mellin transform:
\be \label{MLN}
   m_2\Lb \vec{k}_T,  y,  y_0 \Rb \,\,=\,\,\int\limits^{\epsilon + i  \infty}_{\epsilon - i \infty}\!\!\!\! \frac{d \gamma}{2\,\pi\,i}   e^{ \gamma\,\xi'}  \widetilde{m}_2\Lb \gamma,y,y_0\Rb~~~~~~\mbox{where}~~~\xi'\,\,=\,\,\ln k^2_T
   \ee 
 from both part of \eq{M2EQMOM} we obtain:
 \be \label{M2MLN}
 \frac{\partial\,\widetilde{m}_2\Lb \gamma, y,  y_0 \Rb}{\partial\,y}\,\,=\,\, \chi\Lb \gamma\Rb \widetilde{m}_2\Lb \gamma,  y,  y_0 \Rb\,\,+\,\,\,2\,\widetilde{n^2} 
\ee
where $\chi\Lb \gamma\Rb$ is given by \eq{OMG} and $\widetilde{n^2} $ denotes the Mellin  image of $N^2\Lb \vec{k}_T, \vec{b} , y, y_0 \Rb$. 

\eq{M2MLN} has the following solution which satisfies the initial condition of \eq{M2IC}:

\be \label{SOLMLN}
\widetilde{m}_2\Lb \gamma,  y,  y_0 \Rb\,\,=\,\, e^{\,\chi\Lb \gamma\Rb\,y} \int^y_0\!\!\!d y' \,2\,\widetilde{n^2}e^{\,-\,\chi\Lb \gamma\Rb\,y'}  +\,\,\widetilde{m}^{\rm BFKL}_2\Lb \gamma,  y,  y_0 \Rb
\ee

where $\widetilde{m}^{\rm BFKL}_2\Lb \gamma,  y,  y_0 \Rb$ is a solution to the homogeneous  linear BFKL equation with the initial condition of \eq{M2IC}.  In the following, we neglect the contribution of  this term.~

\subsubsection{Semi-classical solution}


For large $y$ and $\xi$ we can use the semi-classical approximation( SCA,  see Refs.\cite{KOLEB,BKL} and references therein)  to take the integral over $y'$ in \eq{SOLMLN}. In this approximation we are searching for 
\be \label{SCN}
N = e^{S_N} \,=\,e^{ \omega\Lb \xi', y\Rb \,y + \gamma\Lb y,\xi'\Rb \xi'}
\ee
 where $\omega\Lb \xi', y\Rb$ and 
 $ \gamma\Lb y,\xi'\Rb$  are smooth functions of $y$ and $\xi'$:  \,$\partial \,  \omega'\Lb \xi', y\Rb/\partial\,y \,\ll\,\omega^2\Lb \xi', y\Rb, \, \partial \, \omega\Lb \xi', y\Rb/\partial\,\xi'\,\ll\,\omega\Lb \xi', y\Rb\,\gamma\Lb y,\xi'\Rb, \partial \,\gamma\Lb \xi', y\Rb/\partial\, \xi'\,\ll\,\gamma^2\Lb \xi', y\Rb, \, \partial\, \gamma\Lb \xi', y\Rb/\partial \,\xi'\,\ll\,\omega\Lb \xi', y\Rb\,\gamma\Lb y,\xi'\Rb$. Such form  of $N$ stems from \eq{SOLN} if  we use the method of steepest descent for calculating the integral over $\gamma$.  Indeed, using this method one can see that 
 \be \label{MOSD}
  \omega\Lb \xi', y\Rb\,\,=\,\,\chi\Lb  \gamma_{\rm SP} \Rb ;~~~~~\gamma\Lb \xi', y\Rb \,\,=\,\, \gamma_{\rm SP};  ~~~~\mbox{equation for $\gamma_{\rm SP}$:} ~~\frac{d \chi\Lb \gamma\Rb}{d \gamma}\Bigg{|}_{\gamma= \gamma_{\rm SP} }\!\!\!\!\!\!\!\!\!\!\!\!y\,\,=\,\,\xi';
  \ee

  In the SCA the  Mellin image of $N^2$ can be written as follows:
  
  \be \label{N2MLN}
    N^2\Lb \xi',  y,  y_0 \Rb \,\,=\,\,e^{2\,S_N} \,=\,e^{ 2\,\omega\Lb \xi', y\Rb \,y + 2\,\gamma\Lb y,\xi'\Rb \xi'}\,\,=\,\,   \int\limits^{\epsilon + i  \infty}_{\epsilon - i \infty}\!\!\!\! \frac{d \gamma}{2\,\pi\,i}   e^{\,\,2 \chi\Lb\h \gamma\Rb\,y\,\,+\,\, \gamma\,\xi'}  H\Lb \gamma\Rb
   \ee 
  Indeed, taking the integral by the method of steepest descent we obtain the following equation for the saddle point ($\gamma^{(2)}_{SP}$):
  \be \label{MOSD1}
 2\frac{d\, \chi\Lb \h\gamma\Rb}{d \gamma}\Bigg{|}_{\gamma= \gamma^{(2)}_{\rm SP} }y\,\,=\,\,\xi';  \ee
 with the solution $ \gamma^{(2)}_{\rm SP}  \,=\,2\,\gamma_{SP} $, where $\gamma_{SP}$ is given by \eq{MOSD}. Plugging this solution into \eq{N2MLN} we see that 
  \bea \label{N2MLN1}
    N^2\Lb \xi',  y,  y_0 \Rb \,\,&= &\,\, e^{\,\,2 \chi\Lb\h \gamma^{(2)}_{SP}\Rb\,y\,\,+\,\, \gamma^{(2)}_{SP} \,\xi'}\int\limits^{\epsilon + i  \infty}_{\epsilon - i \infty}\!\!\!\! \frac{d \Delta \gamma}{2\,\pi\,i}   e^{\,\,\frac{d^2\chi\Lb \h \gamma^{(2)}_{SP} \Rb}{d \gamma} \,y (\Delta \gamma)^2}  H\Lb  \gamma^{(2)}_{SP} \Rb\nn\\
    &=& e^{\,\,2 \chi\Lb \gamma_{SP}\Rb\,y\,\,+\,\,2\, \gamma_{SP} \,\xi'}\int\limits^{\epsilon + i  \infty}_{\epsilon - i \infty}\!\!\!\! \frac{d \Delta \gamma}{2\,\pi\,i}   e^{\,\,\frac{d^2\chi\Lb \gamma_{SP} \Rb}{d \gamma} \,y (\Delta \gamma)^2}  H\Lb  2 \gamma_{SP} \Rb   \eea 
 The integral over $\Delta \gamma$ leads to a smooth function, which in the SCA can be considered as a constant. Therefore, comparing \eq{N2MLN} and \eq{N2MLN1} one can see that \eq{N2MLN} is correct.  
 We can derive \eq{N2MLN} using a more general consideration.
  Actually, the  expression for the Mellin transform of $    N^2\Lb \xi',  y,  y_0 \Rb $ is the convolution in $\gamma$  of Mellin images of $N$, which has the following form:
\be \label{MLNM21}
\widetilde{n^2}\,\,=\,\,\,\int\limits^{\epsilon + i\,\infty}_{\epsilon - i\,\infty}\!\! \frac{d \,\gamma'}{2\,\pi\,i}n_{in}\Lb \gamma'\Rb\,n_{in}\Lb \gamma\,-\,\gamma'\Rb e^{\Lb  \chi\Lb \gamma'\Rb \,\,+\,\,\chi\Lb\gamma\,-\, \gamma'\Rb\Rb\,y}
\ee
Taking the integral over $\gamma'$ using the method of steepest descent one can see that the equation for the saddle point has the following form:
\be \label{SPGA} 
\frac{ d\,\chi\Lb \gamma  - \gamma'\Rb}{ d\,\gamma'} \,\,+\,\,\frac{ d\,\chi\Lb  \gamma'\Rb}{ d\,\gamma'}\,\,=\,\,0
\ee

with the solution  $\gamma'_{SP} = \h \gamma$. Plugging this solution in \eq{MLNM21} we reduce it to:
\be \label{MLNM22}
\widetilde{n^2}\,\,=\,\,\,e^{ 2 \chi\Lb \h \gamma\Rb\,y}\int\limits^{\epsilon + i\,\infty}_{\epsilon - i\,\infty}\!\! \frac{d \,\Delta \gamma'}{2\,\pi\,i}\,\,n^2_{in}\Lb \h \gamma\Rb e^{\frac{d^2 \chi\Lb \gamma'_{SP}\Rb}{d \gamma'^2_{SP}}\,y\, \Lb \Delta \gamma'\Rb^2}
\ee
which reproduces the  Mellin transform of \eq{N2MLN}.

    Plugging in \eq{SOLMLN} $ \widetilde{n^2}\,\,=\,\,e^{ 2 \chi\Lb \h \gamma\Rb\, y} H\Lb \gamma\Rb$ 
 we can take the integral over $y'$ and the solution has the form:
 
 \be \label{SOLMLN1}
\tilde{m}_2\Lb \gamma,  y,  y_0 \Rb\,\,=\,\,\,e^{\,\,\chi\Lb \gamma\Rb\,y } H\Lb \gamma\Rb \int^y_0\!\!\!d y' \,2\,\widetilde{n^2}\,\,e^{\,-\,\chi\Lb \gamma\Rb\,y'}  
=\,\,\frac{2}{2  \chi\Lb \h \gamma\Rb\,\,-\,\, \chi\Lb \gamma\Rb} \Bigg\{ e^{ 2 \chi\Lb\h \gamma\Rb\, y} H\Lb \gamma\Rb \,\,-\,\, e^{ \chi\Lb \gamma\Rb\, y} H\Lb \gamma\Rb\Bigg\}
\ee
Note, that in \eq{SOLMLN1} we neglected the contribution of  $\widetilde{m}^{\rm BFKL}_2\Lb \gamma,  y,  y_0 \Rb$ in \eq{SOLMLN}.

 Before fixing $H\Lb \gamma\Rb$
  we need to go back to coordinate representation. Indeed,  in this representation we have simple initial conditions for $M_2$  of \eq{M2IC}. Since all solutions are solutions of the linear equations  and $\gamma_{SP}\,\ll\,1$,  we can replace $\xi' =   \ln k^2_T$ by $\xi\,=\,\ln\frac{1}{ r^2}$.
   Bearing this in mind, we can reduce the solution to  the form:
  
 \be \label{SOLMLN2}
M_2\Lb \xi,  y,  y_0 \Rb\,\,=\,\,\int\limits^{\epsilon + i\,\infty}_{\epsilon - i\,\infty}\!\!\!\frac{d\,\gamma}{2\,\pi\,i} \,\frac{2\,\chi\Lb \gamma\Rb}{2\,\chi\Lb \h \gamma\Rb  \,\,-\,\, \chi\Lb \gamma\Rb} \Bigg\{ e^{ 2 \chi\Lb \h \gamma\Rb\, y} \,\,-\,\, e^{ \chi\Lb \gamma\Rb\, y} \Bigg\}\,\frac{1}{\gamma} \,e^{ \gamma\,\,\xi}
\ee
 
 First, we note, that taking the integral over $\gamma$ using the method of steepest descent, we reproduce \eq{SOLMLN1} with the particular choice of $H\Lb \gamma\Rb  =\chi\Lb \gamma\Rb/\gamma$, which  has been  discussed in \eq{NINM}.      Second, one can see that at $ y\,\to\,0$ this solution coincides with \eq{M21I}.
 
 In DLA  this solution takes the form (see \eq{BFKLKER}:
 \begin{subequations}  \bea 
M^{\rm DLA}_2\Lb \xi',  y,  y_0 \Rb\,\,&=&\,\,\int\limits^{\epsilon + i\,\infty}_{\epsilon - i\,\infty}\!\! \frac{d \,\gamma}{2\,\pi\,i}\,\,\frac{2}{3} 
 \Bigg\{ e^{ \frac{4}{\gamma}\, y} \,\,-\,\, e^{ \frac{1}{\gamma}\, y} \Bigg\}\,\frac{1}{\gamma} \,e^{ \gamma\,\,\xi'}\label{SOLMLN31}\\
 &=& 
\frac{2}{3} \Bigg\{    N_{\rm SCA}^2\Lb r,  y, y_0\Rb\,\,-\,\,N_{\rm SCA}\Lb r,  y, y_0\Rb \Bigg\}\,\,\leftarrow \mbox{with semiclassical accuracy}\label{SOLMLN32}
\eea
 \end{subequations}  

 However, it turns out that \eq{SOLMLN31} reproduces at least two terms of the expansion at small values of 
 $y$ of the following relation:
 \be \label{SOLMLN4} 
 M_2\Lb \eq{SOLMLN31}\Rb\,\,=\,\,2\Bigg( N^2\Lb r,  y,y_0\Rb\,\,-\,\,N\Lb r,  y,y_0\Rb\Bigg)\,\,=\,\,
 \,\,2\,y \xi \,\,+\,\,\frac{3}{2} \Lb y\,\xi\Rb^2\,+\,{\cal O}\Lb \Lb y\,\xi\Rb^3\Rb 
  \ee

  Concluding,  we see that  \eq{C1} does not hold in the DLA, but \eq{SOLMLN4} gives us some hope to find an approach in which it will be correct.
  
\subsubsection{Diffusion approximation}


Actually, the most adequate approach  at high energies is the diffusion one (see \eq{BFKLKER}).
Plugging in \eq{SOLMLN2} the BFKL kernel in the form $\chi\Lb \gamma\Rb\,\,=\,\,\omega_0 \,\,+D\,\Lb \gamma\,-\,\h\Rb^2$  and taking the integral using the method of steepest  descent we obtain the saddle point 
\be \label{SP}
\gamma_{SP} \,\,=\,\,\h \,+ i \frac{\xi}{2\,D\,y}; ~~~ \,\,\,2\,\chi\Lb \h \gamma_{SP}\Rb\,\,=\,\,2\,\omega_0\,\,-\frac{\xi^2}{2\,D\,y} ; ~~~~~\chi\Lb  \gamma_{SP}\Rb\,\,=\,\,\,\omega_0\,\,-\frac{\xi^2}{4\,D\,y} ;
\ee
Considering $\frac{\xi}{2\,D\,y}\,\,\ll\,\,1$ we can neglect their contributions in the factor  $\frac{2\,\chi\Lb \gamma\Rb}{2\,\chi\Lb \h \gamma\Rb  \,\,-\,\, \chi\Lb \gamma\Rb}$ reducing it to 2. Hence, \eq{SOLMLN2} reads 
\be \label{SOLMLN6}
M^{\rm DA}_2\Lb r,  y, y_0\Rb\,\,=\,\,2\Bigg( N^2\Lb r,  y, y_0\Rb\,\,-\,\,N\Lb r, y, y_0\Rb\Bigg)
\ee

\begin{figure}[ht]
    \centering
  \leavevmode
      \includegraphics[width=10cm]{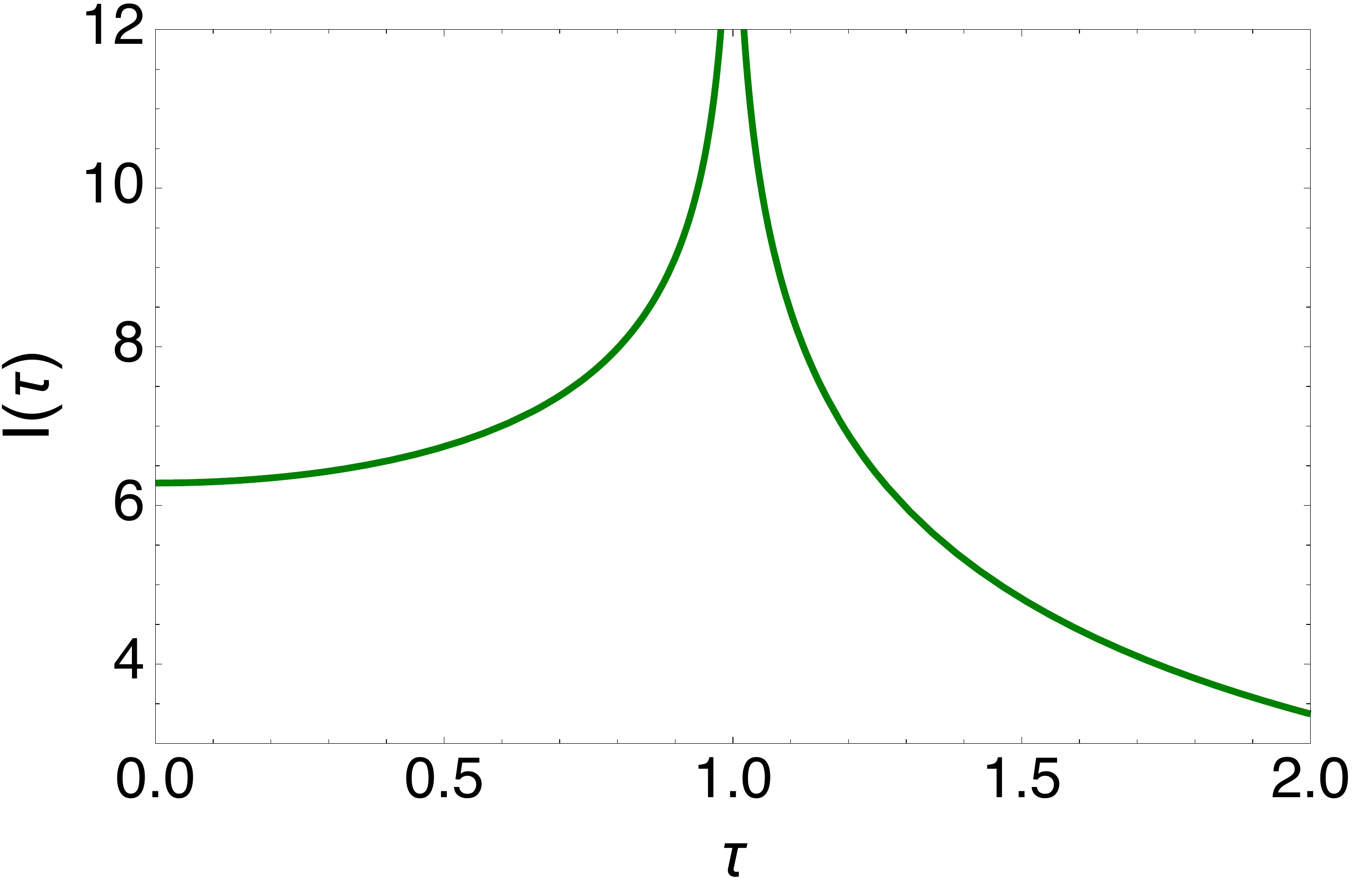}  
      \caption{$I(\tau)$  of \eq{I} versus $\tau=r'/r$. }
\label{ker}
   \end{figure}

We can obtain the solution of \eq{SOLMLN6} directly from the equation for $M_2$  (see \eq{M2EQ}), if we note that the BFKL kernel has   maxima at $r' \to 0$ and $|\vec{r} - \vec{r}'| \to 0$. In \fig{ker} we plot the term of \eq{M2EQ} , which is proportional to $N^2$:
\be \label{I}
 \int d^2 r' K\Lb \vec{r}',\vec{r} - \vec{r'}|\vec{r}\Rb\,2\,N\Lb \vec{r}' , y, y_0 \Rb \,N\Lb \vec{r}\,-\,\vec{r}' , y ,y_0\Rb\,\, \propto\,\,  \int d^2 r'  K\Lb \vec{r}',\vec{r} - \vec{r'}|\vec{r}\Rb\,(\Lb r'^2 \,\Lb \vec{r} - \vec{r'}\Rb^2\Rb^{\h} = \int d r' I\Lb \tau =\frac{r'}{r} \Rb
 \ee
We can see from \fig{ker}, that $I(\tau)$ has a maximum at $\tau = 1$.  Note, that in   \eq{I} we introduce $\gamma = \h$ , which corresponds to the DA, to estimates the  value of  this contribution.
 
 Bearing this observation in mind, we can re-write \eq{M2EQ} in the following form:
  \bea \label{M2EQDA}
\frac{\partial\,M_2\Lb \vec{r}, y,  y_0 \Rb}{\partial\,y}\,\,&=&\,\, \int\limits_{r' < r} d^2 r' K\Lb \vec{r}',\vec{r} - \vec{r'}|\vec{r}\Rb  M_2\Lb \vec{r}, y, y_0 \Rb +  \int\limits_{r' < r} d^2 r'\, K\Lb \vec{r}',\vec{r} - \vec{r'}|\vec{r}\Rb M_2\Lb \vec{r}', y, y_0 \Rb\\
 &+& \int\limits_{|\vec{r} - \vec{r}'| < r}\!\!\!\!\! d^2 r'\, K\Lb \vec{r}',\vec{r} - \vec{r'}|\vec{r}\Rb M_2\Lb \vec{r}\,-\,\vec{r}', y, y_0 \Rb +  \int\limits_{|\vec{r} - \vec{r}'| < r}\!\!\!\!\!  d^2 r' \,K\Lb \vec{r}',\vec{r} - \vec{r'}|\vec{r}\Rb M_2\Lb \vec{r}, y, y_0 \Rb\nn\\
 & \,\,+&\,\,2\, \int\limits_{|\vec{r} - \vec{r}'| < r}\!\!\!\!\! d^2 r' \,K\Lb \vec{r}',\vec{r} - \vec{r'}|\vec{r}\Rb N\Lb \vec{r} , y, y_0 \Rb \,N\Lb \vec{r}\,-\,\vec{r}' , y ,y_0\Rb\nn\\
 & +&
 2 \int\limits_{r' < r} d^2 r'\, K\Lb \vec{r}',\vec{r} - \vec{r'}|\vec{r}\Rb  N\Lb \vec{r} , y, y_0 \Rb \,N\Lb \,\vec{r}' , y ,y_0\Rb 
   -\int d^2 r\, K\Lb \vec{r}',\vec{r} - \vec{r'}|\vec{r}\Rb \,\,M_2\Lb \vec{r},  y, y_0 \Rb\nn
\eea

One can see (1)  that in \eq{M2EQDA} the terms, which  are proportional to $M_2\Lb \vec{r},  y, y_0 \Rb$, cancel each other out, and (2) this equation can be presented in the form:
\be \label{M2EQDA1}
\frac{\partial\,M_2\Lb \vec{r}, y,  y_0 \Rb}{\partial\,y}\,\,=\,\, \int d^2 r' K\Lb \vec{r}',\vec{r} - \vec{r'}|\vec{r}\Rb\Bigg\{M_2\Lb \vec{r}', y, y_0 \Rb \,\,+\,\,2\,N\Lb \vec{r} , y, y_0 \Rb \,N\Lb\,\vec{r}' , y ,y_0\Rb\Bigg\} \ee

Substituting \eq{SOLMLN6} into \eq{M2EQDA1} we reduce it in the SCA to the equation:
\be \label{M2EQDA2}
\frac{\partial\,2\Lb N^2 - N\Rb}{\partial\,y}\,=\, 2\,\chi\Lb \gamma_{SP}\Rb \Lb  \,2 N^2  -  N\Rb\,\,= \,\,2 \Big( \Lb \chi \Lb 2  \gamma_{SP}\Rb\,+\,\chi\Lb \gamma_{SP}\Rb \Rb\,N^2 - \chi\Lb \gamma_{SP}\Rb\,N\Big)
\ee
Hence, one can see,  that in the limit of small $ \frac{\xi}{2\,D\,y}\,\,\ll\,\,1$,
$\chi \Lb 2  \gamma_{SP}\Rb = \chi \Lb   \gamma_{SP}\Rb$ and \eq{SOLMLN6} satisfies  \eq{M2EQDA1} and  \eq{M2EQDA2}.

  It is  worth noting that \eq{SOLMLN4} follows from the multiplicity distribution,  which is given by \eq{MULTD}.
We will concentrate our efforts on DA  in our presentation below.

\subsection{The third moment}

The third moment can be found from the generating function $\tilde{w}_\lambda\Lb \vec{r}, \vec{b}, y, y_0\Rb$ in the following way:

\be \label{M3}
M_3\Lb \vec{r},  y,  y_0 \Rb\,\,=\,\,\,\langle| n (n\,-\,1)\,(n\,-\,2) |\rangle = \sum^\infty_{n=1} n\,(n-1)\,(n-2)w_n\Lb r, y, y_0\Rb\,=\,\frac{\partial^3\,\, w_\lambda\Lb \vec{r},  y, y_0\Rb}{\partial\,\,\lambda^3} \Big{|}_{\lambda\,=\,1}
\ee

 Using \eq{M3}   we obtain the equation for $M_3\Lb \vec{r},  y,  y_0 \Rb$ from \eq{LMMGEN}:
 
 \bea \label{M3EQ}
\frac{\partial\,M_3\Lb \vec{r}, y,  y_0 \Rb}{\partial\,y}\,\,&=&\,\, \int d^2 r' K\Lb \vec{r}',\vec{r} - \vec{r'}|\vec{r}\Rb\Bigg\{M_3\Lb \vec{r}' , y, y_0 \Rb \,+\,M_3\Lb \vec{r}\,-\,\vec{r}', y, y_0 \Rb\nn\\
 & \,\,+&\,\,6\,M_2\Lb \vec{r}' , y, y_0 \Rb \,N\Lb \vec{r}\,-\,\vec{r}', y ,y_0\Rb -\,\,M_3\Lb \vec{r}, y, y_0 \Rb\Bigg\} \nn
\eea
Rewriting \eq{M3EQ} in  momentum representation(see \eq{M2MOM})  we reduce  this eqaution to  the form:
\be \label{M3EQMOM}
\frac{\partial\,m_3\Lb \vec{k}_T,  y,  y_0 \Rb}{\partial\,y}\,\,=\,\, \int \frac{d^2 k'_T}{(2 \pi)^2} K\Lb \vec{k}_T,\vec{k}'_T\Rb\,\,m_3\Lb \vec{k}'_T,  y, y_0 \Rb 
  \,\,+\,\,6\,m_2\Lb \vec{k}_T, y, y_0 \Rb\, n\Lb \vec{k}_T, y, y_0 \Rb   
\ee
where we use lowcase letters denoting the moments in the momentum representation.

The Mellin image of \eq{M3EQMOM}  has the form:
 \be \label{M3MLN}
 \frac{\partial\,\widetilde{m}_3\Lb \gamma,  y,  y_0 \Rb}{\partial\,y}\,\,=\,\, \chi\Lb \gamma\Rb \widetilde{m}_3\Lb \gamma,  y,  y_0 \Rb\,\,+\,\,\,6 \,\widetilde{n^2\,n} 
\ee
where $\widetilde{n^2\,n} $ is the Mellin image of $m_2\Lb \vec{k}_T,  y, y_0 \Rb\, N\Lb \vec{k}_T , y, y_0 \Rb$ which has the general form:

 \be \label{M2MLN70}
\widetilde{n^2\,n}\Lb \gamma, y\Rb\,\,=\,\,\,\int\limits^{\epsilon + i\,\infty}_{\epsilon - i\,\infty} \frac{d \,\gamma'}{2\,\pi\,i}\,
\widetilde{n^2}\Lb \gamma', y\Rb\,\widetilde{n}\Lb \gamma  -  \gamma', y\Rb\ee
where $\widetilde{n^2}\Lb \gamma, y\Rb$ is determined by (see \eq{SOLMLN2})
\be \label{M2MLN7}
\widetilde{n^2}\,\,=\,\,\frac{2\,\chi\Lb \gamma\Rb}{2\,\chi\Lb \h \gamma\Rb  \,\,-\,\, \chi\Lb \gamma\Rb} \Bigg\{ e^{ 2 \chi\Lb \h \gamma\Rb\, y} \,\,-\,\, e^{ \chi\Lb \gamma\Rb\, y} \Bigg\}\,\frac{1}{\gamma}
\ee
Plugging \eq{SOLN} and \eq{M2MLN7} into \eq{M2MLN70}   one can see that $\widetilde{n^2\,n}\Lb \gamma, y\Rb$ is a sun of two terms.  
Taking the integral over $\gamma'$ in each of them in the saddle point approximation  (see \eq{MLNM21} - \eq{MLNM22}) we obtain  two equations for the saddle points:
\be \label{M2MLN71}
2\frac{d\,\chi\Lb \h \gamma'\Rb}{d\gamma' } + \frac{d\,\chi\Lb \gamma -  \gamma'\Rb}{d\gamma' } = 0~~~\mbox{with} ~~~\gamma'_{SP} = \frac{2}{3} \gamma;  ~~~~~
\frac{ d\,\chi\Lb \gamma  - \gamma'\Rb}{ d\,\gamma'} \,\,+\,\,\frac{ d\,\chi\Lb  \gamma'\Rb}{ d\,\gamma'}\,\,=\,\,0;~ ~~\mbox{with}\,\gamma'_{SP} = \h  \gamma;
\ee

Hence,  $\widetilde{n^2\,n}\Lb \gamma, y\Rb$  takes the form:
 \be \label{MLNM211}
\widetilde{n^2\,n}\Lb \gamma, y\Rb\,\,=\,\,\,\frac{2\,\chi\Lb \gamma\Rb}{2\,\chi\Lb \frac{1}{3} \gamma\Rb  \,\,-\,\, \chi\Lb\h \gamma\Rb} \Bigg\{ e^{ 3\chi\Lb \frac{1}{3} \gamma\Rb\, y} \,\,-\,\, e^{ \chi\Lb \h \gamma\Rb\, y} \Bigg\}\,H_2\Lb \gamma\Rb
\ee
The solution does not depend on the  form of $H_2\Lb \gamma\Rb$ , which we will specify below.

Plugging this equation in \eq{M3MLN} we obtain the solution:
\bea \label{SOLM3MLN}
\widetilde{m}_3\Lb \gamma,  y,  y_0 \Rb\,\,&=&\,\,6\Lb\frac{1}{3 \,\chi\Lb \frac{1}{3} \gamma\Rb \,\,-\,\,\chi\Lb  \gamma\Rb}\Rb\,\Lb\frac{2\,\chi\Lb \gamma\Rb}{2\,\chi\Lb \frac{1}{3} \gamma\Rb  \,\,-\,\, \chi\Lb\h \gamma\Rb}\Rb \Bigg\{ e^{ 3\,\chi\Lb \frac{1}{3} \gamma\Rb\,y} \,\,-\,\, e^{ \,\chi\Lb \gamma\Rb\,y}\Bigg\}\,\,\nn\\
 &  -&\,\,\Lb \frac{1}{2\,\chi\Lb \frac{1}{3} \gamma\Rb  \,\,-\,\, \chi\Lb\h \gamma\Rb}\Rb\Lb \frac{2\,\chi\Lb \gamma\Rb}{2\,\chi\Lb \frac{1}{2} \gamma\Rb  \,\,-\,\, \chi\Lb \gamma\Rb}\Rb\Bigg\{ e^{ 2\,\chi\Lb \frac{1}{2} \gamma\Rb\,y} \,\,-\,\, e^{ \,\chi\Lb \gamma\Rb\,y}\Bigg\}\,H_2\Lb \gamma\Rb
 \eea
 Going to the coordinate representation as was discussed above  and choosing $H_2\Lb \gamma\Rb\,=\,\chi\Lb \gamma\Rb/\gamma$, which reproduces the correct initial conditions we obtain 
  \bea \label{SOLM3MLN2}
M_3\Lb \xi, y,  y_0 \Rb\,\,&=&\,\,6\,\int\limits^{\epsilon + i\,\infty}_{\epsilon - i\,\infty}\frac{d\,\gamma}{2\,\pi\,i} ,\Lb\frac{\chi\Lb \gamma\Rb}{3 \,\chi\Lb \frac{1}{3} \gamma\Rb \,\,-\,\,\chi\Lb  \gamma\Rb}\Rb\,\Lb\frac{2\,\chi\Lb \gamma\Rb}{2\,\chi\Lb \frac{1}{3} \gamma\Rb  \,\,-\,\, \chi\Lb\h \gamma\Rb}\Rb \Bigg\{ e^{ 3\,\chi\Lb \frac{1}{3} \gamma\Rb\,y} \,\,-\,\, e^{ \,\chi\Lb \gamma\Rb\,y}\Bigg\}\,\,\nn\\
 &  -&\,\,\Lb \frac{\chi\Lb \gamma\Rb}{2\,\chi\Lb \frac{1}{3} \gamma\Rb  \,\,-\,\, \chi\Lb\h \gamma\Rb}\Rb\Lb \frac{2\,\chi\Lb \gamma\Rb}{2\,\chi\Lb \frac{1}{2} \gamma\Rb  \,\,-\,\, \chi\Lb \gamma\Rb}\Rb\Bigg\{ e^{ 2\,\chi\Lb \frac{1}{2} \gamma\Rb\,y} \,\,-\,\, e^{ \,\chi\Lb \gamma\Rb\,y}\Bigg\}\,\frac{1}{\gamma}e^{ \gamma\,\,\xi}
\eea 

Using the method of steepest descent and neglecting contributions of the order of  $\frac{\xi'}{2 \,D\,y} $  in all  pre-exponential factors,  we see that
\be \label{SOLM3MLN3}
M_3\Lb \xi,  y,  y_0 \Rb\,\,=\,\,6 \,N\Lb \xi, y,  y_0 \Rb\,\Lb N\Lb \xi,  y,  y_0 \Rb\,\,-\,\,1\Rb^2,
\ee
which is the same as for the multiplicity distribution of \eq{MULTD}.

\subsection{General approach}
The equation for a general  moment  
\be \label{MK}
M_k\Lb r,  y, y_0\Rb = \langle| n (n\,-\,1) \dots (n - k+1) |\rangle = \sum^\infty_{n=1}\Lb  n\,(n\,-\,1) \dots (n-k+1)\Rb\,w_n\Lb r,  y, y_0\Rb\,=\,\frac{\partial^k\,\, w_\lambda\Lb \vec{r},  y, y_0\Rb}{\partial\,\,\lambda^k} \Big{|}_{\lambda\,=\,1}
\ee
we can obtain by  differentiating \eq{LMMGEN} with respect to $\lambda$.  It has the simple form in the momentum representation (see \eq{M2MOM}):
\be \label{MKEQ}
\frac{\partial}{\partial\,y}m_k\Lb  k_T, y, y_0\Rb\,\,=\,\,\int \frac{d^2 k_T}{(2 \pi)^2} K\Lb \vec{k}_T,\vec{k}'_T\Rb\,\,m_k\Lb \vec{k}'_T,  y, y_0 \Rb 
  \,\,+\,\,\sum^{k-1}_{i=1}\,\frac{k!}{(k - i)!\,i!}m_{k - i}\Lb \vec{k}_T, \ y, y_0 \Rb\, m_{i}\Lb \vec{k}_T, y, y_0 \Rb  
\ee

As we have discussed, we can go back to coordinate representation  in \eq{MKEQ}, since $m_k$ are solution to the linear equations, and, therefore,  have the Mellin image 

\be \label{MKMLN}
 m_k\Lb \vec{k}_T, y \Rb\,\,\,=\,\,\,\int\limits^{\epsilon \,+\,i\,\infty}_{\epsilon \,-\,i\,\infty}\!\! \frac{d \gamma}{2\,\pi\,i}\,e^{ \omega\Lb \gamma\Rb\,y \,\,+\,\,\gamma\,\xi'}\,m_{k0}\Lb \gamma\Rb~~~~\mbox{with}~~~\xi' \,\,=\,\,\ln k^2_T
 \ee
In coordinate representation we have
\be \label{MKMLN1}
 M_k\Lb r,  y \Rb\,\,\,=\,\,\,\int\limits^{\epsilon \,+\,i\,\infty}_{\epsilon \,-\,i\,\infty}\!\! \frac{d \gamma}{2\,\pi\,i}\,e^{ \omega\Lb \gamma\Rb\,y \,\,+\,\,\gamma\,\xi}\,m_{k0} \Lb \gamma\Rb\,H_k(\gamma)~~~~\mbox{with}~~~\xi \,\,=\,\,\ln\Lb \frac{1}{ r^2}\Rb
 \ee
The coordinate image of 
$m_{k - i}\Lb \vec{k}_T, \ y, y_0 \Rb\, m_{k}\Lb \vec{k}_T, y, y_0 \Rb$ is  $
\int d^2 r'\,K\Lb \vec{r}',\vec{r} - \vec{r'}|\vec{r}\Rb\,M_{k-i}\Lb \vec{r}',  y, y_0 \Rb \,M_i\Lb \vec{r}\,-\,\vec{r}',y, y_0\Rb$. On the other hand, in SCA   $m_{k - i}\Lb \vec{k}_T, \ y, y_0 \Rb\, m_{i}\Lb \vec{k}_T, y, y_0 \Rb $ has the image in $\gamma$-representation, which is equal to ${\rm Const} \exp\Lb k \chi\Lb \frac{1}{k}\gamma\Rb y\Rb\,m_{k-i,0}\Lb \frac{\gamma}{k}\Rb\,m_{i,0}\Lb \frac{\gamma}{k}\Rb\,\widetilde{H}_k\Lb \gamma\Rb$   (see \eq{SOLMLN1} and \eq{MLNM211})\footnote{ Actually, the $\gamma$ image of $m_k$
is a sum of the terms with different $k$ (see \eq{SOLMLN1} and \eq{MLNM211}), but  this does not change  the conclusion.}. Using this image,  one can see, that the coordinate representation for $m_{k - i}\Lb \vec{k}_T, \ y, y_0 \Rb\, m_{k}\Lb \vec{k}_T, y, y_0 \Rb$ can be reduced to $\int d^2 r' K\Lb \vec{r}',\vec{r} - \vec{r'}|\vec{r}\Rb\,\,M_{k - i}\Lb r', y, y_0 \Rb\, M_{i}\Lb r',  y, y_0 \Rb$, which means, that $\widetilde{H}_k\Lb \gamma\Rb\,=\,\chi\Lb \gamma\Rb$ as we expect from \eq{SOLMLN2} and \eq{SOLM3MLN2}.

Finally we can re-write
\eq{MKEQ} in the coordinate representation in the form:
\bea \label{MKEQ2}
\frac{\partial}{\partial\,y} M_k\Lb  r, y, y_0\Rb\,\,&=&\,\, \int d^2 r' K\Lb \vec{r}',\vec{r} - \vec{r'}|\vec{r}\Rb\,
\Bigg\{M_k\Lb \vec{r}',  y, y_0 \Rb \,+\,M_k\Lb \vec{r}\,-\,\vec{r}', y, y_0 \Rb\,-\,M_k\Lb \vec{r} ,y, y_0 \Rb \nn\\
& +&\,\,\,\sum^{k-1}_{i=1}\,\frac{k!}{(k-i)!\,i!}M_{k - i}\Lb r', y, y_0 \Rb\, M_{i}\Lb r',  y, y_0 \Rb\Bigg\} 
\eea
  .

Assuming that for all $i\,\leq\,k - 1$ 
\be \label{MI}
M_i\Lb r , y, y_0 \Rb\,\,=\,\,i! \,N\Lb r, y, y_0 \Rb\,\Lb N\Lb r, y, y_0 \Rb\,-\,1\Rb^{i - 1},
\ee 
 which follows from \eq{MULTD}, we will prove that for $i=k$ we have the same expression.

Plugging \eq{MI} in  \eq{MKEQ2} we get the inhomogeneous term in  the form $k! \Lb k - 1\Rb N^2\Lb N \,-\,1\Rb^{k-2}$. In the following we will use that the Mellin  image of $N^i\Lb r, y, y_0 \Rb$ ($\widetilde{n^i}$) is  equal to
\be \label{MIMLNNI}
\widetilde{n^i}\,\,=\,\,e^{i\,\chi\Lb \frac{1}{i}\gamma\Rb\,y} \frac{1}{\gamma}\,\,=\,\,\,\int\limits^{\epsilon \,+\,i\,\infty}_{\epsilon \,-\,i\,\infty} \frac{d \omega}{2\,\pi\,i}\,e^{\omega \,y} \frac{1}{\omega\,\,-\,\,i\,\chi\Lb \frac{1}{i}\gamma\Rb}\frac{1}{\gamma};~~~~~~~\mbox{double Mellin image} \,\,\widetilde{\widetilde{n^i}}\,\,=\,\,\frac{1}{\omega\,\,-\,\,i\,\chi\Lb \frac{1}{i}\gamma\Rb}\frac{1}{\gamma};\ee
which can be derived using the method of steepest descent  in  the estimates of the integrals over $\gamma$'s (see \eq{MLNM21} for example).

In the double Mellin transform \eq{MKEQ2} takes the form:
\be \label{MIEQ2MLN}
\Lb \omega\,\,-\,\,\chi\Lb \gamma\Rb\Rb\widetilde{\widetilde{M_k}}\,\,=\,\,k!\,(k - 1)\chi\Lb \gamma\Rb\sum^{k-2}_{l=0} 
\frac{(-1)^l (k - 2)!}{(k - 2 - l)! \,l!}\,\frac{1}{\omega\,-\,\Lb k - l\Rb \chi\Lb\frac{1}{k - l}\gamma\Rb}\frac{1}{\gamma}
\ee

Hence from \eq{MIEQ2MLN} we have:
\bea \label{SOLMK}
&&M_k\Lb  r,y, y_0\Rb\,\,=\,\,\,k!\,(k - 1)\sum^{k-2}_{l=0} \frac{(-1)^l (k - 2)!}{(k - 2 - l)! \,l!}\,\\
&& \times\,\,\int\limits^{\epsilon \,+\,i\,\infty}_{\epsilon \,-\,i\,\infty} \frac{d \omega}{2\,\pi\,i}\,\int\limits^{\epsilon \,+\,i\,\infty}_{\epsilon \,-\,i\,\infty} \frac{d \gamma}{2\,\pi\,i}\,e^{\omega \,y\,\,+\,\,\gamma\,\xi'} 
\frac{\chi\Lb \gamma\Rb}{(n - l) \chi\Lb\frac{1}{k - l}\gamma\Rb\,-\,\chi\Lb \gamma\Rb}\Bigg\{
\frac{1}{\omega\,-\,\Lb k - l\Rb \chi\Lb\frac{1}{k - l}\gamma\Rb}\,\,-\,\,\,\frac{1}{\omega\,-\,\chi\Lb \gamma\Rb}\Bigg\}\frac{1}{\gamma}\nn\\
&&=\,k!\,(k - 1)\sum^{n-2}_{l=0} \frac{(-1)^l (k - 2)!}{(k - 2 - l)! \,l!}\int\limits^{\epsilon \,+\,i\,\infty}_{\epsilon \,-\,i\,\infty} \frac{d \gamma}{2\,\pi\,i}\,e^{\,\,\gamma\,\xi'} 
\frac{\chi\Lb \gamma\Rb}{(k - l) \chi\Lb\frac{1}{k - l}\gamma\Rb\,-\,\chi\Lb \gamma\Rb}\Bigg\{
e^{ \Lb  k - l\Rb \chi\Lb\frac{1}{k - 1}\gamma\Rb\,y}\,\,-\,\,\,e^{\chi\Lb \gamma\Rb\,y}\Bigg\}\frac{1}{\gamma}\nn\eea
In \eq{SOLMK} we  calculate  the integrals over $\omega$ closing the contour of integration on  poles.
Taking the integral over $\gamma$ using the method of steepest descent  in the diffusion approximation  and neglecting the corrections of the order of $\xi'/Dy$ (see discussions above)  we reduce \eq{SOLMK}  to the following expression:
\bea \label{SOLMK1}
M_k\Lb  r,  y, y_0\Rb\,\,&=&\,\,k!\,\sum^{k-2}_{l=0} \frac{(-1)^l (k - 1)!}{(k - 1 - l)!\, l!}\Lb N^{k-l}\Lb  r,  y, y_0\Rb\,\,-\,\,N\Lb  r, b, y, y_0\Rb\Rb\,\,\nn\\
&=& k!\,\sum^{k-1}_{l=0} \frac{(-1)^l (k - 1)!}{(k - 1 - l)! \,l!} N^{k-l}\Lb  r,  y, y_0\Rb \,=\,   k!\, N\Lb  r,  y, y_0\Rb\,\Lb N\Lb  r,  y, y_0\Rb\,\,-\,\,1\Rb^{k - 1}
 \eea

Since we have obtained \eq{MI} for $i=2$ and $i=3$ , we prove this equation for any value of $i$.

\section{Finding corrections and comparison with experiments}

\eq{MI} generates the multiplicity distribution of    \eq{MULTD}. However, several questions arise,  when we wish to compare this distribution with the experimental data, let say with DIS. The average multiplicity of the color-singlet dipoles is equal to the sea quark structure function $x\Sigma_{\rm sea}\Lb x, Q^2\Rb$\cite{KHLE1}. On the other hand, in \eq{MI} the multiplicity enters in the coordinate representation.  In diffraction approximation the momentum and coordinate representations are related by the replacement $\ln k^2_T\,\to -\ln r^2$. Therefore, my suggestion is to use \eq{MULTD} with $N\,=x\Sigma_{\rm sea}\Lb x, Q^2\Rb$ but to calculate the corrections to this distribution.  The multiplicity  distribution can generally be written, using the cumulant generating function $f\Lb \lambda\Rb$  as follows \cite{MUMULT,MGF}:
\be
\frac{\sigma_n}{\sigma_{ \rm in}}\,\,=\,\,\oint\frac{d \lambda}{2\,\pi\,i} \frac{e^{ f\Lb \lambda\Rb}}{\lambda^{n + 1}}
\ee
where the contour of  integration  is the circle around the point  $\lambda =0$  and  $f\Lb \lambda\Rb$ is the cumulant generating function, which  is defined as
\be \label{CGF}
f\Lb \lambda\Rb\,\,=\,\,\sum^\infty_{n=1} \frac{\kappa_n }{n!} \Lb \lambda - 1\Rb^n
\ee
where $\kappa_n$ are cumulants. Generally speaking, we have the following definition for the cumulants:
\be \label{CUM}
\kappa_1\,=\,N;~~~\kappa_2\,=\,M_2\,-\,N^2;~~~ \kappa_3\,=\,M_3\,-\,3\,M_2 \,N\,+\,2\,N^2;~~~\kappa_4\,=\,M_4\,-\,4\,M_3\,N\,-\,3\,M^2_2\,+\,12\,M_2\,N^2\,-\,6\,N^4;
\ee
where $M_k$ are the factorial moments, that we have discussed above (see \eq{MI}).

In our case we  can view $f\Lb \lambda\Rb$ as a sum of $f\Lb \lambda\Rb\,\,=\,\,f^{\eq{MULTD}}\Lb \lambda\Rb\,\,+\,\,\Delta f\Lb \lambda\Rb$. $f^{\eq{MULTD}}\Lb \lambda\Rb$ generates the multiplicity distributions of \eq{MULTD}, which includes the most dominant contributions and ,in particular, the average number of color-singlet dipoles is taken into account exactly.  We suggest to introduce function 
$\Delta f\Lb \lambda\Rb$ in the following way:
\be \label{CGFD}
\Delta f\Lb \lambda\Rb\,\,=\,\,\sum^\infty_{n=1} \frac{\Delta\kappa_n }{n!} \Lb \lambda - 1\Rb^n
\ee
with
\be \label{DCUM}
\Delta \kappa_1\,=\,0;~~\Delta \kappa_2\,\,=\,\,M_2\Lb k_T, y\Rb \,-\,2\,N\,\Lb N\,- \,1\Rb ;~~~~
 \Delta \kappa_3\,=\,M_3\Lb k_T, y\Rb\,-  6\,N \Lb  N\,-\,1\Rb^2 \, -\,3\,\Delta \kappa_2 \,N\,;\ee
where $M_2\Lb k_T, y\Rb$  and $M_3\Lb k_T, y\Rb$ are  the exact solutions to \eq{M2EQMOM} and  \eq{M3EQMOM}, respectively.

Introducing two multiplicity distributions:
\be \label{2D}
P_n\Lb N\Rb\,\,=\,\,\,\,\oint\frac{d \lambda}{2\,\pi\,i} \frac{e^{ f^{\eq{MULTD}}\Lb \lambda\Rb}}{\lambda^{n + 1}};~~~~~~~\widetilde{P}_n\Lb N\Rb\,\,=\,\,\,\,\oint\frac{d \lambda}{2\,\pi\,i} \frac{e^{\Delta f\Lb \lambda\Rb}}{\lambda^{n + 1}};
\ee
one can see that the resulting multiplicity distribution takes the form

\be \label{RMULTD}
\frac{\sigma_n}{\sigma_{ \rm in}}\,\,=\,\,\sum^n_{k=0}\,\frac{n!}{(n - k)!\,k!}\,P_{n - k}\Lb N\Rb\, \widetilde{P}_k\Lb N\Rb
\ee

In the case of $\Delta \kappa_2\,\neq\,0$ but  $\Delta \kappa_n\,=\,0$ for $n\,>\,2$ the distribution $\widetilde{P}_n\Lb N\Rb$ has been found in Ref.\cite{MUMULT} and it has the following form in our notations:

\be \label{TILDEP}
\widetilde{P}_n\Lb \Delta \kappa_2\Rb \,\,=\,\,e^{ \frac{\Delta \kappa_2}{2}}\frac{\Lb - i \sqrt{ \frac{\Delta \kappa_2}{2}}\Rb^n}{n!}\,{\rm H_n}\Lb i \sqrt{ \frac{\Delta \kappa_2}{2}}\Rb
\ee
where ${\rm H}_n $ is   Hermite polynomial (see formula {\bf 8.95} in Ref.\cite{RY}).

\section{Conclusions}
 
This paper has two main 
results. 
First, we derived the BFKL linear, inhomogeneous equation for the factorial moments of  multiplicity distribution ($M_k$) from LMM  equation. In particular, the equation for the average multiplicity of the color-singlet ($N$)
turns out to be the homogeneous  BFKL equation which leads to the power-like growth in the region of small $x$.  From these equations it follows that $M_k\,\,\propto\,\,N^k$ at small $x$.

Second, using the diffusion approximation for the BFKL kernel, which is generally  considered  to be responsible for the small $x$  behaviour, we show that the factorial moments  satisfy \eq{C1},
 which  reproduces the multiplicity distribution of \eq{MULTD}. This result is in agreement with the attempts \cite{GOLE}   to find solutions to the equations for the cascade of  color-singlet dipoles (see \eq{PC1}.
 
 We also suggest a procedure for finding corrections to this multiplicity distribution, which, we believe, will be useful for descriptions of the experimental data.
 
 In general, the multiplicity distribution, that has been discussed in the paper,  confirms the result of Ref.\cite{KHLE}, that the entropy  of color-singlet dipoles  is equal $S\,\,=\,\,\ln N$ in the region of small $x$, and gives the regular procedure to estimate corrections to this formula.
 
 It is worthwhile mentioning that both SCA and DA have been developed before and many technical issues that matter,  have been discussed\footnote{We thank our referee who drew our attention to Ref.\cite{BWX},  where  some technical details of our approach have been clarified.}(see, for example,  Refs.\cite{KOLEB,LRSH,LLSH,BRAUN1,BWX,LEMI} and references therein).

\section{Acknowledgements}
   We thank our colleagues at Tel Aviv university and UTFSM for
 encouraging discussions.  Our special thanks go to Michael Lublinsky  for fruitful discussion of Ref.\cite{LMM}. 
This research was supported  by 
 ANID PIA/APOYO AFB180002 (Chile) and  Fondecyt (Chile) grants  
 1180118.

\end{document}